\documentstyle[12pt]{article}
\begin{document}
\baselineskip = 20pt

\title{QUANTUM CORRECTIONS TO THE TWO DIMENSIONAL GRAVITY WITH EXTERNAL
FIELD}

\author{ M.Alves$^{\star}$\\
\\
Theory Division - CERN \\
CH-1211 GENEVE 23 \\
Switzerland}

\bigskip
\maketitle
\begin{abstract}
We introduce an external field to calculate the quantum corrections to the
solutions of the 2d gravity, via trace anomaly. We show that there are
black hole type solutions even in the absence of matter field and cosmological
constant. We also see that these solutions are similar to the ones
obtained from dilaton two dimensional gravity.
\end{abstract}
\bigskip
\bigskip
\vfill
\bigskip
\vfill
\noindent $^\star${e-mail: ALVES@VXCERN.CERN.CH}

\noindent On leave from
\noindent Instituto de Fisica - Universidade Federal do Rio de Janeiro-BRASIL
\par

\pagebreak
\noindent
{\bf 1- INTRODUCTION}
\par Relativistic theories of gravitation in two spacetime dimensions have been
studied intensively from long time, the main motivation being the possibility
of obtaining relevant information about issues of the classical and
quantum
relativistic theory of gravitation in four spacetime dimension.
Recent work has  shown that these two dimensional models have a rich and
interesting structure: gravitational collapse, cosmological models and
gravitational
waves are some of the aspects that can be easily studied within this framework.
Besides this, it has become apparent that these theories have also remarkable
implications to conformal field theories and string motived effective
models.
\par By two dimension gravity (2d gravity) we mean a relativistic theory of
the metric tensor in two dimensions (1+1) spacetime. Some of these models have
been studied by several authors [1, 2 and references therein] and, in
particular, two of
them have
 features that concern us. In [1], the authors begin with an action with
 metric and dilaton fields coupled with matter, called dilaton gravity.
The model is solved  and with cosmological constant it
shown to have black hole solutions even in the absence of matter sources. It is
possible to identify the mass of this black hole
and when it is null, another
black hole type solution is found, in this case, with a matter source included.

\par Another point of view consist in looking for an analogue to the Einstein
equation. However, since in two dimensions the Einstein-Hilbert action is a
topological invariant, we
can not extract any dynamics from it. In despite of this, non-trivial
structures
arise from the 2d version of the Einstein equation [2,3]:

\begin{equation}
R - \Lambda = 8\pi G T ,
\end{equation}

\noindent where R, $\Lambda$, T and G are the Ricci scalar, the cosmological
constant, the trace of the energy-momentum tensor (EMT) and the gravitational
constant respectively. Taking a point particle as the matter source, there are
non-trivial structures [2], and
black hole solutions arise even when $\Lambda=0$ [2,3].

\par It is in this framework it seems easier to take into account
quantum corrections as one could adopt the semiclassical version of (1), given
by

\begin{equation}
R - \Lambda = 8\pi G <T> ,
\end{equation}

\noindent
where quantum effects are given by the vacuun expectation value (VEV)
of the trace of EMT whereas, the metric tensor is left classical. This equation
gives us the one loop correction.

\par This semiclassical equation means interaction between quantum
fields, via VEV, and a classical one, the metric. So, differently as expected
from a fully quantized theory, anomalies can arise. The trace anomaly is an
example of this effect: a massless field has a null trace of the classical
EMT, a signal of the conformal invariance of the action. When quantum effects
are taken into account semiclassically, the VEV of the trace
of the EMT
is different from zero [4]. This lack of invariance can be understood as
being due to
the interaction with the classical field, the metric. Incidentally, we mention
that even in the flat spacetime case the interaction between classical and
quantum fields produces this anomaly [5].

\par So, one could expect that the trace anomaly
introduces modifications to the Einstein equation via (2), yielding the lower
order quantum corrections.
An example of this approach is the massless field case:
since the trace of the classical EMT is zero, the corrections in (2) are due
entirely to the conformal anomaly. The right hand side of (2) being
different from zero, would give us other solutions than those obtained
from the classical version. However, this is not the case. Both bosonic and
fermionic massless fields in two dimensions have the associated anomaly
proportional to $R$, the Ricci scalar, [4] and we can  redefine the
gravitational constant in a such way that the semiclassical equation turns
to be the same as
the classical one without any new information.

\par The main purpose of this note is to modify this scenario. As mentioned
before, the trace anomaly is
an effect of the interaction between classical and quantum fields. When others
fields are considered the anomaly must be different from the previous
value obtained when considering only the gravitational field as the classical
interacting field. In [8] it has been used an auxiliary external classical
field
coupled to the bosonic field and shown that the trace anomaly change to
a more general expression. Here we make use of this new expression for the
anomaly and calculate the corrections to the classical solutions of (1).
Moreover, as these solutions are the same black hole type solutions found
in [1],
from the dilaton gravity, we can speculate that they are closely
related. An interpretation of the external field as a kind of non-minimal
coupling with gravity is also given.
This paper is organized as follows: we begin with the motivation for
introducing the auxiliary external
field, then we construct the modified two dimension Einstein equation and
calculate the solutions for some few cases. In the conclusion, we discuss
the  possibility of these corrections to be raised from a
new coupling between the scalar and the gravitational fields.

\bigskip
\bigskip
\bigskip
\noindent
{\bf 2 - THE TRACE ANOMALY WITH AN EXTERNAL CLASSICAL FIELD}
\par In constructing a relativistic theory for the two dimensional scalar
field in curved spacetime, some modifications have to be introduced, namelly
a suitable redefinition for the field variable, described as follows.
\par For the action to this theory one can write

\begin{equation}
S=\int d^4x \sqrt{-g} g^{\mu\nu}\partial_\mu\phi\partial_\nu\phi ,
\end{equation}

\noindent the EMT associated is given by

\begin{equation}
T_{\mu\nu}={2\over \sqrt{-g}}{\delta S\over \delta g^{\mu\nu}}
=\partial_\mu\phi\partial_\nu\phi - {1\over 2}g_{\mu\nu}\partial_{\alpha}\phi
\partial^{\alpha}\phi
\end{equation}

\noindent and the equation of motion

\begin{equation}
\nabla^{\mu}\partial_{\mu}\phi = {1\over \sqrt{-g}}\partial_{\mu}(\sqrt{-g}
\partial^{\mu}\phi)=0 .
\end{equation}

\par The EMT is conserved, which means

\begin{equation}
\partial^{\nu}T_{\mu\nu}=0 .
\end{equation}

\par The semiclassical approach makes use of the VEV of these quantities and
some requirements must be satisfied [4], one of them being the conservation of
the VEV of the EMT. However, whereas the trace of the classical EMT is
automatically zero (only in  two dimensions), its conservation is valid only
on-shell via the equation of motion (5):

\begin{equation}
\langle T^{\mu}_{\mu}\rangle=g^{\mu\nu}\langle T_{\mu\nu}\rangle=
\langle\partial^{\mu}\phi\partial_{\mu}\phi\rangle-\langle\partial_{\alpha}\phi
\partial^{\alpha}\phi\rangle=0 ,
\end{equation}

\begin{equation}
\partial^{\nu}\langle T_{\mu\nu}\rangle =
\langle\partial_{\mu}\phi \nabla^\nu\partial_\nu\phi\rangle .
\end{equation}

\noindent
The condition (5) can not be used in the semiclassical version (8) and the
requirement on the conservation of the VEV of EMT (8) is not observed.

\par To circunvent this undesirable result, we can change the field
variable [6] to

\begin{equation}
\tilde\phi=g^{1/4}\phi
\end{equation}

\noindent
and now, it is straightforward to see that considering as independent
variables the pair  $ g_{\mu\nu}$ and $\tilde\phi$ , we obtain an
expression for the EMT
that, after rewrited  again in terms of $g_{\mu\nu}$ and $\phi$, is

\begin{equation}
\tilde T_{\mu\nu}=\partial_\mu\phi\partial_\nu\phi .
\end{equation}

Now, the trace of the EMT is zero only on-shell.
In this way, the trace anomaly arises, as expected

\begin{equation}
\langle \tilde T^\mu_\mu \rangle=
\langle\phi \nabla^\mu\partial_\mu\phi\rangle= {1\over 24\pi}R
\end{equation}

\noindent and the (regularized) expression to the conservation of the EMT is
the usual one, as it should be, [6]

\begin{equation}
\nabla^{\mu}\langle \tilde T_{\mu\nu} \rangle = 0 .
\end{equation}

\par We can find the origin of this change of the field variable on
fundamental
grounds: it is necessary to make the full quantized theory free of a
non-invariant BRST jacobian, which renders the theory free of anomalies [7].
\par The new variable, differently from the old one, is not conformally
invariant under a conformal transformation of the metric

\begin{equation}
g^{,}_{\mu\nu}=e^{2\alpha(x)}g_{\mu\nu} ,
\end{equation}

\begin{equation}
\tilde\phi^{,}=e^{\alpha(x)}\tilde\phi
\end{equation}

\noindent
and, consequently, so is not the derivative. In [8], we have introduced an
auxiliary external field, $A_\mu$, to construct a conformally
covariant derivative

\begin{equation}
D_{\mu}\tilde\phi=(\partial_{\mu}+A_{\mu})\tilde\phi ,
\end{equation}

\noindent
with the suitable transformation for the field $A_{\mu}$.
Using this conformal gauge field, we can construct a conformally invariant
action for the (new) field variable

\begin{equation}
S_{inv}=\int d^{2}x\, g^{\mu\nu} D_{\mu}\tilde\phi D_{\nu}\tilde\phi .
\end{equation}

\par The invariance of (16) is ensured by the transformation of the
gauge field (the absence of the factor $\sqrt{-g}$ is due to the fact
that we have now density quantities).

\par The new EMT is

\begin{equation}
T_{\mu\nu}={1 \over \sqrt{-g}}\biggl\{ \partial_{\mu}\tilde\phi
\partial_{\nu}\tilde\phi + \tilde\phi^{2}A_{\mu}A_{\nu}
+ {1\over 2}\tilde\phi^{2}(\partial_{\mu}A_{\nu}
+\partial_{\nu}A_{\mu})\biggr\}
\end{equation}

and the trace

\begin{equation}
T^{\mu}_{\mu}=-\tilde\phi(\nabla^{\mu}\partial_{\mu}-A^{\mu}A_{\mu}
+\partial^{\mu}A_{\mu})\tilde\phi
\end{equation}

vanishes using the equation of motion.
\par The trace anomaly for this theory is given by [8]

\begin{equation}
\langle T^{\mu}_{\mu}\rangle=
{1\over 24\pi}(R-A^{\mu}A_{\mu}+\partial^{\mu}A_{\mu}) .
\end{equation}

We finish this section noting that it is possible to write the
field $A_{\mu}$ in terms of the metric. We can re-write the original action
(3), using $\phi=g^{{-}{1\over 4}}\tilde\phi$, as

\begin{equation}
S=\int d^{2}x \, \sqrt{-g}g^{\mu\nu}\partial_\mu({\tilde\phi\over g^{1/4}})
\partial_{\nu}({\tilde\phi\over g^{1/4}})
\end{equation}

\noindent
or

\begin{equation}
S=\int d^{2}x \, \biggl\{\partial^\mu\tilde\phi \partial_\mu\tilde\phi +
g^{1/4}\partial^{\mu}({1\over g^{1/4}})g^{1/4}\partial_{\mu}({1\over g^{1/4}})
\tilde\phi^{2}+ 2\tilde\phi\partial^{\mu}\tilde\phi
g^{1/4} \partial_{\mu}({1\over g^{1/4}})\biggr\} .
\end{equation}

Comparing with (16)  we have the explicit expression for the
conformal gauge field

\begin{equation}
A_{\mu}=g^{1/4}\partial_{\mu}({1\over g^{1/4}}) .
\end{equation}

The conclusion is  that this field is giving us a new coupling
between the correct field variable and gravity. This kind of non minimal
coupling is similar to that introduced in the 4 dimensional case by the
requirement of
the conformal invariance and generates a different theory. We will show in the
next
section how this  two dimensional  non-minimal coupling changes
the solutions of (1).

\bigskip
\bigskip
\bigskip
{\bf 3 - QUANTUM CORRECTIONS WITH EXTERNAL FIELD}
\par One of the most important features of these models we are considering here
is that there are solutions that can be interpreted as black hole solution
[1,2,9].
We will solve (2) with the modifications introduced in the previous section
for the same
particular cases used in [2]. As a matter of simplicity and in order to
compare results,
we consider the static case and the metric given by

\begin{equation}
g_{\mu\nu}= e^{\sigma}\eta_{\mu\nu}
\end{equation}

\noindent with

\begin{equation}
\sigma=\sigma(x) .
\end{equation}

\par As mentioned before, we will discuss the model given by (2), with the
RHS being only the trace anomaly

\begin{equation}
\langle T \rangle = R + a( \partial^{\mu}A_{\mu} -
A^{\mu}A_{\mu} ) .
\end{equation}

\par In consequence of (23) and (24), the relevant geometrical quantities
are

\begin{equation}
R_{\mu\nu\sigma\tau}=-{1\over 2}(g_{\mu\sigma}g_{\nu\tau}
-g_{\nu\sigma}g_{\mu\tau})R ,
\end{equation}

\begin{equation}
R_{\mu\nu}={1\over 2} g_{\mu\nu} R ,
\end{equation}

\begin{equation}
R= - e^{-\sigma}{d^{2}\sigma \over dx^{2}}
\end{equation}

and, the conformal gauge field (22),

\begin{equation}
A_{\mu}=-{1\over 2} {d\sigma \over dx^{\mu}} .
\end{equation}

\par The term with $\partial^{\mu}A_{\mu}$ is proportional to $R$ and can be
absorbed in another redefinition of constants. The same does not occurs with
$A^{\mu}A_{\mu}$ that contributes with

\begin{equation}
A^{\mu}A_{\mu}={1\over 4}e^{-\sigma}\left({d\sigma \over dx}\right)^{2} .
\end{equation}

\par Using these expressions in (2) and (25) with another suitable
redefinition
of the constants, we have the following equation of motion:

\begin{equation}
e^{-\sigma}{d^{2}\sigma \over dx^{2}}
-ae^{-\sigma}\left({d\sigma \over dx}\right)^{2}
= - \Lambda .
\end{equation}

\par This equation has a more general form than the usual one [2] due to the
presence of the conformal gauge field which is due to the new coupling
with the gravity. With $a=0$, which means the theory without anomaly, (30)
becomes a particular case of the Liouville equation and is one of the models
proposed for 2d gravity [3]. The presence of the non linear term is
due to quantum effects.
Let us now analyse some of its solutions:

\bigskip
\bigskip

\noindent
i) $\Lambda=0$ and $a=0$ (without anomaly)
\par This is the trivial case, since from (31)

\begin{equation}
- e^{-\sigma}{d^{2}\sigma \over dx^{2}}= R = 0
\end{equation}

\noindent
and by (26) the spacetime is then  flat.

\bigskip
\bigskip

\noindent
ii) $\Lambda=0$ and $a \not=0$ (with anomaly)
\par We have now an effective contribution from the anomaly, giving the
quantum corrections.
This correction changes the above equation to

\begin{equation}
e^{-\sigma}{d^{2}\sigma \over dx^{2}}
- ae^{-\sigma}\left( {d\sigma \over dx}\right)^{2}=0 .
\end{equation}

It is easy to see that a solution is

\begin{equation}
e^{-\sigma}=Ax-C
\end{equation}

with A and C constants of integration and we have made $a=1$

Substituting this solution in to the expressions (28) and (26) we see that

\begin{equation}
R_{\mu\nu\lambda\tau} \sim  R \sim  {A^{2} \over (Ax-C)^3}
\end{equation}

\noindent
which is  a non-trivial spacetime. Furthermore, this is the same
black hole solution found
in [2] without cosmological constant, but here we do not have mass as a source
of matter.
This is a new result: In a massless theory without quantum corrections,
the absence of
the cosmological constant we have only the trivial solution, like
in case i). Besides this, we can identify the mass of the black hole as
proportional to the constant $ A $.

\bigskip
\bigskip
\noindent
iii) $\Lambda \not= 0$ and $a \not= 0$
\par The resulting equation is

\begin{equation}
e^{-\sigma}{d^{2}\sigma \over dx^{2}}
- e^{-\sigma}\left({d\sigma \over dx}\right)^{2} = - \Lambda
\end{equation}

\noindent
with  a solution as (again with $a=1$)

\begin{equation}
e^{-\sigma}=-{1 \over 2}\Lambda x^{2} + Bx + C
\end{equation}

\noindent
and again we have non trivial spacetime solutions. Taking $B=0$, (37) is
the same black hole type solution of the dilaton gravity (in the static case)
with $ C = M \sqrt{{2 \over \Lambda}}$, where $M$ is the mass of
the black hole [1], with a suitable choice of signals of the constants.
We mention also that this solution
reproduces that one
from [2], for the $\Lambda \not=0$ case. The novel feature is that, like ii),
it was not necessary a source of matter to generate a non-trivial spacetime.

\bigskip
\bigskip
\bigskip
\noindent
{\bf 4 - CONCLUSIONS AND FINAL REMARKS}
\par In the preceeding section we have found some solutions  of the modified
equation for the
two dimensional gravity.
 The main conclusion is that
when the correct field variable is used, the quantum corrections
are given from the trace anomaly with the external field. This field is
introduced by the
requirement of conformal invariance of the action and the
corrections now can be interpreted as
the usual way, namely
as being due to the fact that the trace anomaly gives a length scale
to the theory. This is the
origin of the "mass" of the black hole in cases ii) and iii).
We emphasize that without the
new variable and, consequently, without $A_{\mu}$ the theory
would be "transparent" to
quantum effects ( at least those given by the trace anomaly). This
fact and the expression for the field in terms of the metric (22)
show  the role of the
conformal gauge field: it creates another coupling between gravity
and the quantum field (in our
case, the scalar one). So, we do not need to construct the dynamics to this
field as it is already
given by the left hand side of (1). Incidentally, we mention that the
conformal covariant
derivative (15) works as  the covariant one for the density field $\tilde\phi$.

\par Another interesting result of the quantum corrections given
by the anomaly can be noted in the case ii) where we have a black hole
without either mass or cosmological constant in the equation of motion.
Whereas the  mass of the black hole is created by the anomaly, as
discussed above, in the
dilaton gravity there is no such a solution without a cosmological constant.
But we have shown showed that the theory
given by the modified 2d Einstein equation (2) has  black holes  in the
absence of this constant. So, the possibility of the modified equation (2)
to be a general case for  both models must be taking in to account.

\par We can also use the conformal gauge field as an auxiliary field to give
the equation of
motion (1) as  derived from an action, but this does  not give any new
consequence. In fact,
the same is done in [9] with a field without dynamics interpreted as a
Lagrange multiplier.
Anyway, all the cases discussed there are reproduced.

\par Finally we mention the the use of  these ideas to some 2d
cosmological models must be investigated.
In despite of that all this analysis have been done to a particular,
static, case we hope
that others features can be obtained when time dependence is allowed.
This is under investigation.

\bigskip
\bigskip
\bigskip
\vfill\eject

\bigskip

\noindent
{\bf REFERENCES}

\noindent
[1] C.G.Callan, S.B.Giddings, J.A.Harvey, A.Strominger, PRD 45, 4,
R1005 (1992); For a recent review, see A.Strominger, "Lectures on black holes",
Les Houches (1994).

\noindent
[2] R.Mann, A.Shiekm, L.Tarasov, Nucl.Phys.B, 341, 134 (1990)

\noindent
[3] R.Jackiw in Quantum Theory of Gravity ed S.Christensen (Adam Hilger,
Bristol, 1984)

\noindent
\,\,C.Teietlboim in Quantum Theory of Gravity, ed S.Christensen (Adam Hilger,
Bristol, 1984)

\noindent
[4] N.D.Birrel, P.C.W.Davies in Quantum Fields in Curved Spacetime (Cambridge
University Press, Cambridge, 1984)

\noindent
[5] M.Alves, J.Barcelos-Neto, Class.Quantum Grav.5, 377 (1988)

\noindent
[6] K.Fujikawa, U.Lindstrom, N.K.Rocek, P.van Nieuwenhuizen, PRD 37, 391 (1988)

\noindent
[7] K.Fujikawa in Quantum Gravity and Cosmology ed H.Sato and T.Inami (World
Scientific, Singapore, 1982)

\noindent
[8] M.Alves, C.Farina, Class.Quantum Grav.9, 1841 (1992)

\noindent
[9] R.Mann, T.G.Steele, Class.Quantum Grav.9, 475 (1992)

\end{document}